# Effect of vacancies on the mechanical properties of zirconium: An *ab initio* investigation

Xueyan Zhu [1,2], Xingyu Gao [1,2,3], Haifeng Song [1,2,3], Guomin Han [1,2] and De-Ye Lin [1,2,3,*]

[1] Institute of Applied Physics and Computational Mathematics, Fenghao East Road 2, Beijing 100094, P.R. China

[2] CAEP Software Center for High Performance Numerical Simulation, Huayuan Road 6, Beijing 100088, P.R. China

[3] Laboratory of Computational Physics, Huayuan Road 6, Beijing 100088, P.R. China

**Abstract**

It is well known that the irradiation-induced defects strongly influence the mechanical properties of zirconium (Zr) or its alloys in nuclear reactors. However, how the point defect changes the mechanical properties has been rarely studied. Here, we systematically investigated the effect of vacancies on the mechanical properties of $\alpha$-Zr based on density functional theory (DFT). Both uniformly distributed vacancies and vacancy clusters were considered. Our results reveal that the existence of vacancy will reduce the bulk modulus, while enhance the shear and Young's moduli. Based on these moduli, the ductility and hardness were further calculated. With the introduction of vacancy, the ductility decreases, but the hardness increases. However, when the vacancy concentration is larger than a critical value, a rise in the ductility and a reduction in the hardness occur, which indicates the degeneration of the material. Moreover, it was found that the vacancies lead to a more isotropic distribution of Young's modulus in 3D space. To further investigate how the clustering of vacancies influences the mechanical properties, the most stable configurations of di- and trivacancy clusters have been predicted, which correspond to the most compact distribution of vacancies. Compared with the uniform distribution of vacancies, clustering of vacancies will strengthen the above changes of elastic moduli, ductility and hardness. Our results indicate that the increase of the vacancy concentration may be the basic cause for the changes of the mechanical properties under irradiation, while the formation of vacancy clusters intensifies these changes.

---

* Corresponding author: De-Ye Lin, email address: lin_deye@iapcm.ac.cn

## 1. Introduction

Zirconium (Zr) and its alloys are widely used as the fuel claddings in nuclear reactors owing to their low capture cross section to thermal neutron, good mechanical properties, and high corrosion resistance in water environment. While in service, they are subjected to a fast neutron flux, leading to the collisions between the neutrons and the atoms. This process will produce high concentration of point defects, including vacancies and interstitials. Then the point defects diffuse and aggregate to form larger defects, such as the dislocation loops and cavities [1, 2]. These irradiation-induced defects have significant impact on the mechanical properties of Zr [3].

Extensive experimental studies [4-11] have confirmed that neutron irradiation increases the strength and hardness, while decreases the ductility as functions of neutron fluence and irradiation temperature. The changes of these mechanical properties have been widely explained from the aspects of the evolutions of defect microstructures during irradiation. For example, the irradiation hardening is attributed to the formation of $\langle a \rangle$ dislocation loops on the prism planes ($\{10\bar{1}0\}$) [7, 12, 13], which act as the obstacles against the dislocation glide [7]. It should be noted that formation of defect microstructures results from the condensation of vacancies or self-interstitials [2, 7]. Therefore, understanding how the point defect influences the mechanical behaviors should be of more basic concern for revealing the fundamental mechanisms of the irradiation-induced changes of the mechanical properties. For such a study, *ab initio* calculations appear to be a suitable tool, which can provide accurate predictions of mechanical properties on the atomic scale that is not accessible by experiments [14].

Recently, several density functional theory (DFT) calculations have been conducted to investigate the effects of vacancy on the mechanical properties of Zr. Zheng *et al.* [15] calculated the elastic constants of Zr with a single vacancy. They found an increase in $C_{11}$, $C_{33}$, $C_{44}$ and a decrease in $C_{12}$ and $C_{13}$. Olsson *et al.* [16] reported a rise in the unstable stacking fault energy and a lowering of the surface energy due to the presence of vacancy, which implies a reduction in ductility. Although several conclusions have been drawn, these calculations were only for single vacancy, and the effects of vacancy concentration and its distribution were not considered. In addition, there were no systematic discussions of the mechanical properties.

The work of this manuscript was dedicated to a systematic investigation of the effect of vacancies on the mechanical properties of α-Zr through DFT calculations. Zr with mono-, di-, and trivacancies were considered. The configurations of these vacancy clusters were determined in the first place. Then the elastic constants, the mechanical stability, stiffness, ductility, hardness, and elastic anisotropy were calculated. The variations of these mechanical properties with respect to the vacancy concentration were obtained and discussed for both uniformly distributed vacancies and vacancy clusters.

This paper is organized as follows. The computational model and methods used in this work were described in section 2. In section 3.1, the stability of di- and trivacancy clusters were investigated. Then the effects of vacancies on the mechanical properties were shown and discussed in section 3.2. Finally, the paper ends with a short summary of the conclusions in section 4.

## 2. Model and Methods

Our DFT calculations were performed using the Vienna Ab-initio Simulation Package (VASP) [17, 18]. The electron-electron exchange-correlation energy was calculated through the generalized gradient approximation (GGA) as parameterized by Perdew, Burke and Ernzerhof (PBE) [19]. The $4s^24p^64d^25s^2$ orbitals of Zr were included as valence electrons and solved by Kohn-Sham (KS) equation [20], while the electron-ion interaction was described by the projector augmented wave (PAW) approach [21, 22]. Plane wave basis set was used with the energy cutoff of 410 eV after convergence tests. The Brillouin zone integrations were performed with a gamma-centered k-point mesh of density of 0.18 $Å^{-1}$. To perform the integration over eigenvalues, the Methfessel-Paxton function of order 1 with a smearing width of 0.2 eV was used as the smearing technique [23]. This smearing ensures the convergence of the difference between the free energy and the total energy within 1 meV/atom. The electron self-consistent iteration was stopped when the total energy change is smaller than $10^{-5}$ eV. The geometry of the system was optimized using the conjugate-gradient (CG) algorithm until all the forces acting on the atoms were smaller than 0.01 eV/Å. Based on the optimized geometry, a static calculation was performed using the tetrahedron method with Blöchl corrections to obtain the accurate energy [24].

The calculations of this paper were performed according to the following procedures. Firstly, the lattice constants and mechanical properties of perfect Zr were calculated and compared to the experimental values to check the reliability of the settings. Then, the same procedure was employed for Zr with vacancies of different concentrations. Both uniformly distributed vacancies and vacancy clusters were considered. The vacancy formation energy and interaction energy have been calculated for determining the most stable configurations of the vacancy clusters.

The lattice constants of α-Zr were obtained by relaxing the primitive unit cell. As shown in Table 1, the calculated lattice constants of α-Zr are in good agreement with experiments and previous calculations. The simulation boxes used for the following calculations were constructed based on the optimized primitive unit cell.

To model the uniformly distributed vacancies, we used the simulation boxes with 2×2×2, 3×3×2, 3×3×3, 4×4×2, 5×5×2, 5×5×4 unit cells containing one vacancy, respectively. The box is periodic in all three dimensions. Thus, the vacancy concentration ranges from 0.005 to 0.063. We modeled di- and trivacancy clusters in the simulation boxes with 5×5×4 unit cells. The geometry of the simulation box and the positions of the atoms were all optimized.

**Table 1**

Lattice parameters (*a*, *c/a*), bulk modulus (*B*) and elastic constants ($C_{ij}$) of α-Zr calculated from DFT calculations, compared with experimental values and other DFT results from the literatures.

| Ref. | *a* (Å) | *c/a* | *B* (GPa) | $C_{11}$ (GPa) | $C_{12}$ (GPa) | $C_{13}$ (GPa) | $C_{33}$ (GPa) | $C_{44}$ (GPa) |
|---|---|---|---|---|---|---|---|---|
| This work | 3.233 | 1.606 | 94.0 | 141.7 | 72.6 | 63.9 | 162.1 | 23.6 |
| Exp [25] | 3.230 | 1.593 | 95.2 | 143.4 | 72.8 | 65.3 | 164.8 | 32.0 |
| DFT-PW91 [26, 27] | 3.230 | 1.606 | 95.3 | 145.3 | 67.3 | 69.5 | 166.1 | 24.3 |
| DFT-PBE [28] | 3.236 | 1.597 | 96 | 146.7 | 68.5 | 71.0 | 163.3 | 26.0 |
| DFT-PBE [29] | 3.238 | 1.600 | 93.6 | 143.5 | 65.9 | – | 168.5 | 25.5 |

Vacancy formation energy describes the energy required for the formation of vacancies and was calculated by

$$E_n^{\text{f}} = E(N-n) - \frac{N-n}{N} E(N),\qquad(1)$$

where *N* is the total number of atoms for the perfect lattice, *n* the number of vacancies, *E*(*N−n*) the total energy of lattice with *n* vacancies, and *E*(*N*) the total energy of the perfect lattice. To characterize the interaction between the vacancies, we further calculated the binding energy of the vacancies, which is defined as the difference between the formation energy of the isolated vacancies and that of the cluster

$$E_n^{\text{b}} = n E_1^{\text{f}} - E_n^{\text{f}}.\qquad(2)$$

Positive $E_n^{\text{b}}$ indicates that the interaction between the vacancies in the cluster is attractive and the configuration of the cluster is stable.

The elastic constants were calculated by the second derivative of the energy with respect to the strain

$$C_{ij} = \frac{1}{V_0}\left(\frac{\partial^2 E}{\partial \varepsilon_i \partial \varepsilon_j}\right),\qquad(3)$$

where $V_0$ is the equilibrium volume, *E* the total energy and $\varepsilon$ the strain in Voigt notation. Due to the hexagonal symmetry, α-Zr has five independent elastic constants ($C_{11}$, $C_{12}$, $C_{13}$, $C_{33}$, $C_{44}$). To obtain these elastic constants, we applied five independent deformations to the equilibrium lattice: (ε, ε, ε, 0, 0, 0), (0.5ε, 0.5ε, -ε, 0, 0, 0), (0, 0, ε, 0, 0, 0), (0, ε, 0, 0, 0, 0), (0, 0, ε, 2ε, 0, 0). ε ranges from -0.01 to 0.01. These five strains result in five independent energy-strain relationships, which were fitted to a polynomial function. From the second-order term of the polynomial function, the elastic constants were extracted. The obtained elastic constants of α-Zr were summarized in Table 1, which showed good agreement with experiments and previous calculations.

Based on the elastic constants of single crystal, the isotropic elastic moduli of polycrystal can be calculated using the Voigt [30], Reuss [31], and Hill [32] approximations, respectively. The Voigt approximation gives the upper bound of the elastic moduli, while the Reuss approximation gives the lower bound. The Hill approximation is the arithmetic average of the upper and lower bounds. For hexagonal crystal, the bulk modulus $B$ and shear modulus $G$ given by Voigt, Reuss, and Hill approximations are

$$
\begin{aligned}
B_{\mathrm{V}} &= \frac{2\left(C_{11}+C_{12}\right)+4C_{13}+C_{33}}{9} \\
G_{\mathrm{V}} &= \frac{1}{30}\left(C_{11}+C_{12}+2C_{33}-4C_{13}+12C_{44}+12C_{66}\right)
\end{aligned}
\tag{4}
$$

$$
\begin{aligned}
B_{\mathrm{R}} &= \frac{C_{33}\left(C_{11}+C_{12}\right)-2C_{13}^{2}}{C_{11}+C_{12}+2C_{33}-4C_{13}} \\
G_{\mathrm{R}} &= \frac{5}{2}\frac{C_{44}C_{66}\left[C_{33}\left(C_{11}+C_{12}\right)-2C_{13}^{2}\right]}{3B_{\mathrm{V}}C_{44}C_{66}+\left(C_{44}+C_{66}\right)\left[C_{33}\left(C_{11}+C_{12}\right)-2C_{13}^{2}\right]}
\end{aligned}
\tag{5}
$$

$$
\begin{aligned}
B_{\mathrm{H}} &= \frac{B_{\mathrm{V}}+B_{\mathrm{R}}}{2} \\
G_{\mathrm{H}} &= \frac{G_{\mathrm{V}}+G_{\mathrm{R}}}{2}
\end{aligned}
\tag{6}
$$

Through the bulk and shear moduli, the Young's modulus $E$ and Possion's ratio $\nu$ can be obtained by

$$
E = \frac{9BG}{3B+G},
\tag{7}
$$

and

$$
\nu = \frac{3B-2G}{2\left(3B+G\right)}.
\tag{8}
$$

## 3. Results and discussion

### 3.1. Configurations of small vacancy clusters

At first, the formation energy of single vacancy $E_1^{\mathrm{f}}$ was calculated through Eqs. (1). The calculated $E_1^{\mathrm{f}}$ is 1.920 eV, which is consistent with experimental values of $E_1^{\mathrm{f}} \geq 1.5\,\mathrm{eV}$ by positron annihilation spectroscopy (PAS) [33, 34] and previous DFT calculations (1.86 – 2.07 eV) [16, 35-38].

Neutron irradiation on Zr will produce more than one vacancy. Therefore, understanding how the vacancy interacts and determining the configurations of the vacancy clusters should be a prerequisite before exploring the effect of vacancies on the mechanical properties or the evolutions of the defect microstructures under radiation. In this manuscript, we only investigated the properties of di- and trivacancies, which are the bases for establishing larger vacancy clusters.

**Table 2**

Binding energy $E_n^b$ for divacancies and trivacancies with different configurations. $d$ is the distance between two vacancies. $\chi$ is a dimensionless number defined by Eqs. (9), which characterizes the compactness of the vacancy cluster. The purple, orange and blue balls represent Zr atoms lying in the $z = 0$, $z = c/2$ and $z = c$, respectively. The purple ball was enlarged for making it visible. The vacancies were marked by the black squares.

| Divacancy | | | | Trivacancy | | | |
|---|---|---|---|---|---|---|---|
| **Index** | **Configuration** | $d$ (Å) | $E_2^b$ (eV) | **Index** | **Configuration** | $\chi$ | $E_3^b$ (eV) |
| **a** |  | 3.199 | 0.148 | **a** |  | 1.745 | 0.408 |
| **b** |  | 3.235 | 0.069 | **b** |  | 1.732 | 0.371 |
| **c** |  | 4.549 | -0.131 | **c** |  | 1.732 | 0.263 |
| **d** |  | 5.195 | -0.266 | **d** |  | 1.433 | 0.251 |
| **e** |  | 5.602 | -0.064 | **e** |  | 1.538 | 0.216 |
| **f** |  | 5.582 | -0.058 | **f** |  | 1.438 | 0.158 |
| **g** |  | 6.119 | -0.057 | **g** |  | 1.500 | 0.091 |
| | | | | **h** |  | 1.473 | 0.091 |
| | | | | **i** |  | 1.431 | 0.054 |

The configurations of the divacancies were chosen such that the distance $d$ between two vacancies is less than the eighth nearest neighbors of Zr as shown in Table 2. The variation of the binding energy $E_2^b$ with respect to $d$ was plotted in Fig. 1(a). $E_2^b$ first decreases from being positive (configurations a and b) to negative (configurations c-g), then increases towards zero as $d$ increases. This indicates that the interaction between two vacancies is attractive only when the two vacancies are within the second nearest neighbors. The most stable configuration of divacancy lies along $(1/6)\left[2\bar{2}03\right]$ direction (configuration a). When $d$ is larger than or equal to the third nearest neighbors, the two vacancies start to repel each other and the divacancy becomes unstable. The most unstable configuration of divacancy corresponds to the fourth nearest neighbors and lies along $\left[0001\right]$ direction (configuration d). For divacancy with $d$ larger than the fourth nearest neighbors, the repulsive interaction becomes weaker. From the above discussions, we can conclude that two vacancies tend to aggregate.

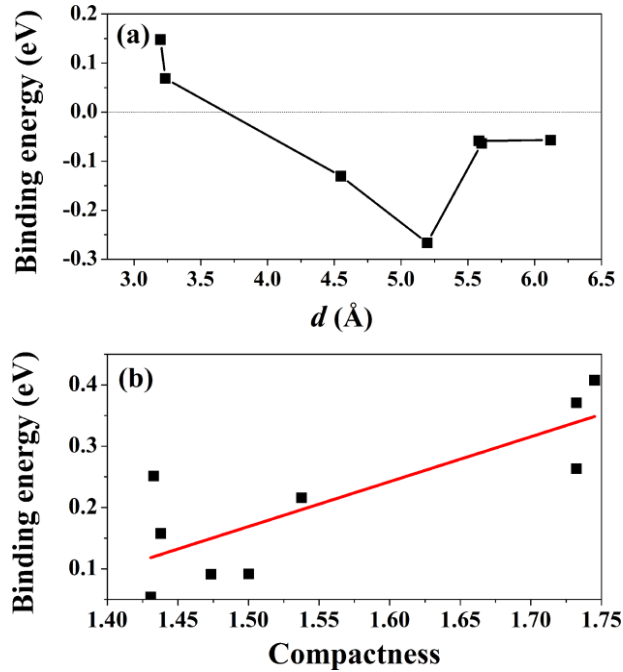

**Fig. 1.** (a) Variation of the binding energy of a divacancy with respect to the distance $d$ between two vacancies. (b) Variation of the binding energy of a trivacancy with respect to the *compactness* $\chi$ defined by Eqs. (9). Black points are obtained from DFT calculations, and the red line is a linear fit to these points.

The stability of trivacancy was further investigated. Based on the above results, only the configurations which contain divacancy within the second nearest neighbors were chosen for calculations as shown in Table 2. All the calculated binding energies are positive, which indicates that the configurations of these trivacancies are all stable. To characterize the clustering degree of the vacancy clusters, we defined a dimensionless number $\chi$ called *compactness* as the ratio of the lattice constant *a* to the average distance between the cluster center and the vacancies *r*:

$$\chi = a \, / \, r \, . \tag{9}$$

Larger $\chi$ indicates higher degree of clustering. As shown in Fig. 1(b), the binding energy increases with $\chi$ although some scattered points were found. The scattered points in Fig. 1(b) indicate that *compactness* is not the only factor that influences the binding energy. Other factors, such as the local atomic arrangements and the stacking topology of vacancies, can also affect the binding energy. For example, the *compactness* of configurations b and c is the same, but their binding energies are different. This is attributed to the different local atomic arrangements, although they have the same stacking topology. Nevertheless, the most stable trivacancy (configuration a) corresponds to the largest $\chi$. Therefore, we can conclude that three vacancies tend to aggregate to form the most compact distributions.

Actually, the aggregation of di- and trivacancies has been experimentally verified by positron annihilation [39]. DFT calculations by Varvenne *et al.* [37] also indicate the stability of di- and trivacancy clusters.

However, the most stable trivacancy predicted by Varvenne *et al.* corresponds to configuration b in Table 2. This may be due to the different strategies used for the lattice relaxations and energy calculations. They kept the volume constant during relaxations and applied an elastic correction to calculate the energy, while we relaxed both the atomic positions and the simulation box, and calculated the energy without any corrections.

### 3.2. Mechanical properties of Zr with vacancies

Two factors that could influence the mechanical properties were considered. One is the distribution of the vacancies, and another is the vacancy concentration. For the distribution of vacancies, we modeled two limiting cases: the uniformly distributed vacancies and the most stable vacancy clusters predicted in section 3.1. The concentration of the uniformly distributed vacancies was controlled by the size of the simulation box, while the concentration of the vacancy cluster by the number of the constituent vacancies. The modelling details have been described in section 2.

The calculated elastic constants of Zr with vacancies of different distributions and concentrations were summarized in Table 3. Through these elastic constants, the mechanical stability, stiffness, ductility, hardness, and elastic anisotropy were further derived. In the following discussions, we would first explore the variations of these mechanical properties with respect to the vacancy concentration for Zr with uniformly distributed vacancies, then consider the effect of vacancy clustering on these properties.

For hexagonal crystals, the Born elastic stability conditions [40, 41] are given by:

$$C_{44} > 0,$$
$$C_{11} > |C_{12}|, \tag{10}$$
$$(C_{11} + 2C_{12})C_{33} > 2C_{13}^2.$$

All of the calculated elastic constants in Table 3 satisfy the above conditions. This indicates that the existence of the vacancy does not destroy the mechanical stability of Zr.

**Table 3**
Elastic constants $C_{ij}$ (in GPa) of Zr with vacancies of different distributions and concentrations. $N_v$ is the number of vacancies in the calculated system. $c_v$ is the vacancy concentration.

| Vacancy distribution | Simulation box | $N_v$ | $c_v$ | $C_{11}$ | $C_{12}$ | $C_{13}$ | $C_{33}$ | $C_{44}$ |
|---|---|---|---|---|---|---|---|---|
| | 2×2×2 | 1 | 0.063 | 135.9 | 55.7 | 58.9 | 147.4 | 27.4 |
| | 3×3×2 | 1 | 0.028 | 147.8 | 49.5 | 64.2 | 158.5 | 30.8 |
| Uniformly distributed | 3×3×3 | 1 | 0.019 | 148.4 | 51.9 | 66.2 | 162.2 | 26.9 |
| vacancy | 4×4×2 | 1 | 0.016 | 147.4 | 59.4 | 64.9 | 154.9 | 28.0 |
| | 5×5×2 | 1 | 0.010 | 146.6 | 58.4 | 65.2 | 170.6 | 24.1 |
| | 5×5×4 | 1 | 0.005 | 147.1 | 62.5 | 65.7 | 164.6 | 23.8 |
| Perfect lattice | 1×1×1 | 0 | 0.000 | 141.7 | 72.6 | 63.9 | 162.1 | 23.6 |
| Vacancy cluster | 5×5×4 | 2 | 0.010 | 147.6 | 58.7 | 64.5 | 165.8 | 25.3 |
| | 5×5×4 | 3 | 0.015 | 149.3 | 51.3 | 61.4 | 164.6 | 26.6 |

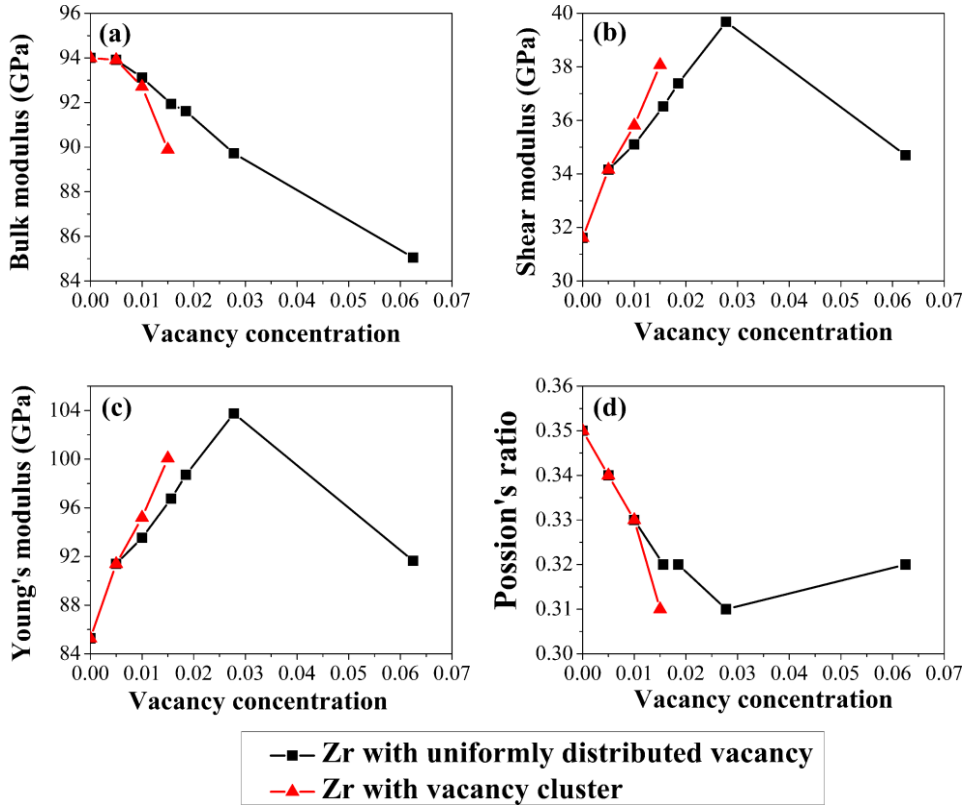

**Fig. 2.** Variations of the bulk, shear, Young's moduli and Possion's ratio with respect to the vacancy concentration, respectively. The black line is for Zr with uniformly distributed vacancy, while the red line for Zr with vacancy cluster.

In the actual industrial applications, Zr is commonly prepared in the polycrystalline form. To examine the effect of vacancies on the stiffness of the polycrystalline Zr, the isotropic elastic moduli have been calculated by Eqs. (4-8). Only the calculations based on Hill approximations were used for the following discussions.

The variations of the bulk, shear, Young's moduli and Possion's ratio with respect to the vacancy concentration $c_v$ were displayed in Fig. 2. With the increase of $c_v$, the bulk modulus $B$ decreases monotonically, which implies the reduction of polycrystalline Zr's resistance to uniform compression. However, the shear modulus $G$, Young's modulus $E$, and Possion's ratio $\nu$ do not change monotonically with $c_v$. For $c_v \leq 0.03$, $G$ and $E$ increase with $c_v$, while $\nu$ decreases. These indicate that the existence of vacancy with low concentration could enhance polycrystalline Zr's resistance to shear stress, uniaxial stress, and transverse contraction under tensile deformation. When $c_v > 0.03$, the reduction of $G$ and $E$, and the rise of $\nu$ occur. This indicates the degeneration of the material when the vacancy concentration is large enough.

The above discussions were focused on the influences of vacancy on the elastic properties. However, irradiation damage could not only lead to the elastic deformation, but also to the plastic deformation. Therefore, we further probed the vacancy effects on the ductility and hardness.

The plastic properties of pure polycrystalline metals have been related to the elastic properties by Pugh [42] in 1954. And he proposed the Pugh's ratio for characterizing the ductility

$$k = G/B . \tag{11}$$

The material is ductile for $k < 0.5$, while brittle for $k > 0.5$. With a larger $k$, the material will behave in a less ductile way. As shown in Fig. 3(a), all the $k$ values are smaller than 0.5. And $k$ values for $c_v > 0$ are all larger than that for $c_v = 0$. This indicates that although Zr with vacancies is still ductile, but should deform in a less ductile way compared with perfect Zr.

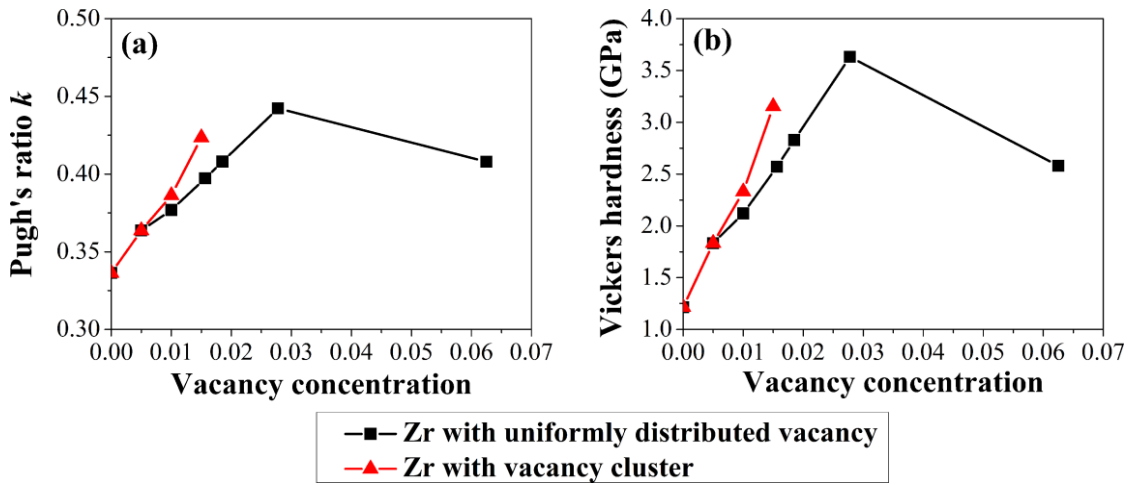

**Fig. 3.** Variations of (a) Pugh's ratio and (b) Vickers hardness with respect to the vacancy concentration. Increase of Pugh's ratio indicates the decrease of the ductility.

The hardness can also be calculated by $B$ and $G$. According to Chen's model [43], the Vickers hardness can be obtained by

$$H_{\mathrm{V}} = 2\left(k^2 G\right)^{0.585} - 3 , \qquad (12)$$

where the units of $H_{\mathrm{V}}$ and $G$ are GPa. Although Chen *et al.* emphasized that their model may be not accurate to predict the hardness of pure metals, the Vickers hardness of Zr predicted by their model is 1.21 GPa, which is in good agreement with experimental values of 0.82 – 1.80 GPa [44]. Therefore, Chen's model was used for exploring the impact of vacancy on the hardness of Zr.

As shown in Fig. 3(b), the Vickers hardness for $c_{\mathrm{v}} > 0$ is larger than that for $c_{\mathrm{v}} = 0$. This indicates the hardening of Zr due to the existence of vacancies. With increasing vacancy concentration, the hardness first increases, then decreases. This variation trend is in agreement with the variation of the hardness with respect to the neutron fluence by the experimental observations of Cockeram *et al.* [9].

According to Cockeram *et al.*, the hardness increase is attributed to the increase of the number density of ⟨a⟩ loops, while the hardness decrease due to the coarsening of ⟨a⟩ loops. Our results indicate that the changes of the hardness under irradiation are not only controlled by the evolutions of the defect microstructures, but the increase of the vacancy concentration also contributes to the changes.

In section 3.1, we have concluded that two and three vacancies tend to aggregate to form clusters. The red lines in Fig. 2-3 show the effect of vacancy clusters on the mechanical properties. The general trend of the changes is the same as that for the uniformly distributed vacancies. However, the clustering of vacancies obviously enhances these changes. And this enhancement becomes more obvious with increasing vacancy concentration. This indicates that vacancy clustering is a factor that exacerbates the changes of the mechanical properties under irradiation.

Furthermore, the effect of vacancies on the elastic anisotropy of single crystal Zr was explored, which plays a key role in the microstructure evolutions of polycrystal [45]. Young's modulus as a function of direction in 3D space for Zr with different vacancy concentrations was plotted in Fig. 4. For the perfect lattice, Young's modulus along $c$ direction is larger than that along $a$ direction. With the increase of vacancy concentration, Young's modulus along $a$ direction gradually increases, while that along $c$ direction does not change much until $c_{\mathrm{v}} = 0.019$. This leads to the more isotropic distribution of Young's modulus. When $c_{\mathrm{v}}$ is increased to 0.063, Young's moduli along $a$ and $c$ directions both decrease, and the energy distribution range becomes narrow. These phenomena indicate the further decrease of the anisotropy. Therefore, we can conclude that the introduction of vacancy will decrease the anisotropy of Young's modulus.

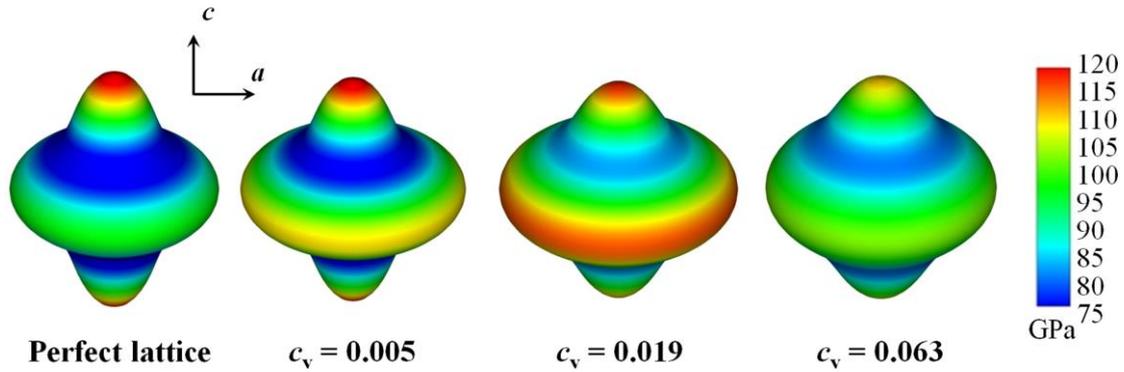

**Fig. 4.** The distribution of Young's modulus as a function of direction in 3D space for perfect Zr and defective Zr with different vacancy concentrations $c_v$.

## 4. Conclusions

Employing DFT calculations, the properties of vacancy clusters and the effect of vacancies on the mechanical properties of Zr have been systematically investigated. It was found that the vacancies tend to aggregate to form vacancy clusters. And the most stable di- and trivacancy clusters correspond to the most compact distribution of vacancies.

In general, the introduction of vacancies decreases the bulk modulus, but increases the shear and Young's moduli. For the vacancy concentration ranging from 0.5% to 6.25%, the maximum changes of the bulk, shear, and Young's moduli are 9.52%, 25.53%, and 21.68%, respectively. Based on these moduli, the ductility and hardness were further calculated. The presence of vacancies contribute to a reduction in the ductility, while a rise in the hardness. With the continuous increase of the vacancy concentration, the bulk modulus monotonically decreases. However, the shear modulus, Young's modulus, and hardness first increase to a maximum value, then decrease. Inversely, the ductility decreases to a minimum value followed by a slight increase. These variations indicate that, when the vacancy concentration is large enough, the degeneration of Zr occurs. Moreover, the elastic anisotropy was explored. It was found that the distribution of Young's modulus becomes more isotropic due to the introduction of vacancies. Compared with the uniform distribution of vacancies, the clustering of vacancies will strengthen the above changes of elastic moduli, ductility and hardness.

In summary, the results of this work indicate that the increased vacancy concentration should be another important factor that contributes to the irradiation-induced changes of the mechanical properties in addition to the evolutions of the defect microstructures. And the clustering of the vacancies enhances these changes.

## Acknowledgements

The authors acknowledge Han Wang for careful reading of the manuscript and giving valuable suggestions. This work was jointly supported by the National High Technology Research and Development Program of China (Grant No. 2015AA01A304) and the National

Natural Science Foundation of China (Grant No. 61300012 and 11176002). The *ab initio* calculations were carried out on the Tianhe-2 clusters supported by the National Supercomputer Centre in Guangzhou.